\documentstyle[aps,epsfig]{revtex}
\begin{document}
\title{The deconfinement phase transition, hadronization \\ and the NJL model}
\author{Sibaji Raha$^*$}
\address{$^*$Physics Department, Bose Institute\\
93/1, Acharya Prafulla Chandra Road\\
Calcutta 700 009, INDIA\\
Electronic Mail : sibaji@bosemain.boseinst.ernet.in}
\maketitle

\begin{abstract}
One of the confident predictions of QCD is that at sufficiently high
temperature and/or density, hadronic matter should undergo a thermodynamic
phase transition to a colour deconfined state of matter - popularly called
the Quark-Gluon Plasma (QGP). In low energy effective theories of Quantum
Chromodynamics (QCD), one usually talks of the chiral transition for which
a well defined order parameter exists. We investigate the dissociation of
pions and kaons in a medium of hot quark matter decsribed by the Nambu -
Jona Lasinio (NJL) model. The decay widths of pion and kaon are found to
be large but finite at temperature much higher than the critical temperature
for the chiral (or deconfinement) transition, the kaon decay width being much
larger. Thus pions and even kaons (with a lower density compared to pions)
may coexist with quarks and gluons at such high temperatures. On the basis
of such premises, we investigate the process of hadronization in quark-gluon
plasma with special emphasis on whether such processes shed any light on
acceptable low energy effective theories of QCD.
\end{abstract}

Quantum Chromodynamics (QCD) is believed to be {\it the underlying
theory} of strong interactions. Although enormously successful, inasmuch
as it is part of the standard model of physical interactions, the
applicability of QCD is studying nuclear interactions or hadronic processes
at low energies is still limited. The difficulties are largely technical,
associated with the non-perturbative features of the theory and hence, a
large amount of effort is being devoted to finding effective lagrangians
which contain features compatible with the low energy limit of QCD. This is
the thrust of the present workshop. The philosophy of my talk is somewhat
complementary; I am going to discuss QCD phenomena which occur in very
energetic nuclear collisions, where one hopes perturbative QCD does play an
important role, at least during the initial time period. 
\par
A confident prediction of QCD is that at very high temperature and/or
density, the bulk properties of strongly interacting matter would be governed
by the coloured QCD degrees of freedom - quarks and gluons- rather than the
usual hadrons. Such a phase is called quark gluon plasma (QGP) \cite{a} in
the literature. Conditions conducive to the formation of QGP may have
existed in the early universe during the first few microseconds after the
Big Bang. Also, such conditions may be transiently created in highly
energetic collsions of large nuclei and the search for
such a novel phase of matter constitutes a major area of current research 
in the field of high energy physics.
\par
Whether QGP is separated from the hadronic world by an actual thermodynamic
phase transition is an open question. It has been widely postulated that
such a phase transition may indeed occur, where the quarks and gluons
convert into colourless hadrons. Recent results, showing the lack of
thermodynamic equilibrium \cite{g} in the quark-gluon phase in
ultrareletivistic heavy ion collisions, indicate however that such an ideal
situation is unlikely. It should also be noted that although the persistence
of non-perturbative effects till very high
temperatures was suggested in the literature quite early on \cite{h}, it is
only recently that the lattice results have confirmed that non-perturbative
hadron like excitations could survive at temperatures far above the chiral
phase transition temperature \cite{i}. The lattice result for pion screening
mass has been studied in ref. \cite{j}. The analysis of \cite{j} has been
contradicted by Boyd {\it et al.} in ref. \cite{ja}. The conclusion of these 
authors \cite{ja} is consistent with the existence of free quarks 
at high temperatures. On the other hand, Shuryak \cite{jb} argued in a
subsequent work that the non-perturbative modes, especially pion- like
excitations, could indeed survive till temperatures above $T_{c}$.
Furthermore, similar results for pion screening 
masses are obtained in $\sigma$- model as well \cite{ja}.
It is thus imperative to understand the behaviour of
such hadronic resonances, their formation, stability and so on, in a
quark gluon medium at high temperature. We confine
our attention to the case of pions and kaons only; these, being lighter  
than other hadrons, account for the bulk of the multiplicity. 
\par
Formation of light mesons like pions and kaons, a bound state of light 
relativistic quarks, is an extremely difficult problem to handle in QCD,
where all the troublesome features of non-perturbative QCD appear.
We therefore employ the usual practice of looking at the pion and kaon as
Goldstone bosons arising from the spontaneous breaking of the chiral
symmetry, a feature most suitably addressed in the Nambu Jona-Lasinio (NJL)
model \cite{k}.  

The formulation of NJL model in flavour SU(3) was first introduced by
Hatsuda {\it et al.} \cite{m} and Bernard {\it et al.} \cite{n}.
The three flavour NJL model Lagrangian is written in terms of $u$, $d$ and
$s$ quarks, the interaction between them being constrained by the
$SU(3)_{L} \otimes SU(3)_{R}$ chiral symmetry, explicit symmetry breaking 
due to the current quark masses and the $U(1)_{A}$ breaking due to the axial 
anamoly \cite{m}. The full Lagrangian with KMT (Kobayashi- Maskawa
-'t-Hooft) term is given below \cite{m}.
\begin{eqnarray}
\cal {L} &=& \bar{q}(i\gamma\cdot\partial - {\bf m})q 
+ {1\over 2}g_{s} {\sum_{a=0}}^{8} [(\bar{q}\lambda_{a}q)^{2} + 
(\bar{q}i\lambda_{a}\gamma_{5}q)^{2}] \nonumber \\
&+& g_{D}[det \bar{q_i}(1-\gamma_{5})q_{j} + h.c.]
\label{eq:lnjl}
\end{eqnarray}
where the quark fields $q_{i}$ has three colours ($N_{c}=3$) and three
flavours ($N_{f}=3$) and $\lambda_{a}$ ($a= 1,8$) are the Gell-Mann 
matrices. The quark mass matrix is given by 
${\bf m}= diag(m_{u},m_{d},m_{s})$.

In the mean field approximation, the quark condensates at finite temperature
are given by,
\begin{eqnarray}
<<\bar{q_{i}} q_{i}>>= - 2N_{c}\int {d^3p \over{2 \pi^3}} 
{M_{i} \over {E_{ip}}} f(E_{ip})
\label{eq:conden}
\end{eqnarray}
where $E_{ip}$ is the quark single particle energy for the i-th specie and 
$f(E_{ip})=1 - n_{ip} - \bar{n}_{ip}$, $n_{ip}$ and $\bar {n}_{ip}$ being 
the Fermi-Dirac distributions for quarks and anti-quarks, respectively. 
If quark chemical potential is zero, then 
$n_{ip}= \bar {n}_{ip} = [exp(E_{ip}/T)~~+~~1]^{-1}$.

The temperature dependent constituent quark masses $M_{i}$ are obtained
from the expressions below,
\begin{eqnarray}
M_{u}=m_{u} - 2g_{s}\alpha - 2g_{D}\beta\gamma   \nonumber \\
M_{d}=m_{d} - 2g_{s}\beta - 2g_{D}\alpha\gamma   \nonumber \\ 
M_{s}=m_{s} - 2g_{s}\gamma - 2g_{D}\alpha\beta    
\label{eq:conmas}
\end{eqnarray}
where
\begin{eqnarray}
<<\bar{u} u>> \equiv \alpha,~~~~<<\bar{d} d>> \equiv \beta, 
~~~~<<\bar{s} s>> \equiv \gamma
\label{eq:defalpha}
\end{eqnarray}

We now want to investigate the decay of pionic and kaonic excitations, the
properties of which we assume to be given by the NJL model. It should be
mentioned here that at temperatures above the critical temperature, these
mesonic excitations are more like resonances with large effective masses
\cite{i,j}. In the following, we study \cite{mpla} the decay width of such
pseudoscalar excitations in the hot quark medium as a function of temperature,
starting with the Lagrangian given above in equation (\ref{eq:lnjl}). 
\par
The quark mass $M_{i}$ appearing in eq. (\ref{eq:conden}) and 
in eq. (\ref{eq:conmas}) is a very important
ingredient in our calculation. In the absence of any medium and/or
dynamic effect, $M_{i}$ is the current quark mass. 
On the other hand, we know that due to the spontaneous breakdown of the
chiral symmetry, quarks attain the value of the constituent quark mass 
\cite{nn}.
\par
In the present calculation we have used the parametrisation of 
ref. \cite{n} ( $\Lambda$ = 631.4, $g_{s}\Lambda^2$ = 3.67, 
$g_{D}\Lambda^5$ = -9.29 and 
current mass $m_{u,d}(m_s)$ = 5.5 (135.7) MeV ) to calculate the quark and 
meson masses. The constituent quark masses are calculated using the 
gap equations(eq. \ref{eq:conmas}). These quark masses are then put into 
the dispersion equation  for mesons to get dynamical masses of 
mesons ($\pi$ and $K$, here).
\begin{eqnarray}
1~~ +~~ 2G_{\phi}\Pi_{ij}(\omega,\vec{q}\rightarrow 0)~~ =~~ 0
\end{eqnarray}
where $\Pi_{ij}$ is the one loop polarization due to $u$ and $d$ quark for 
pions and $u$ or $d$ and $s$ quark for kaons. $G_{\phi}$ is the coupling 
constant with $\phi$ coerresponding to $\pi$ or $K$. 
The general expression for polarization function is 
\begin{eqnarray}
\Pi(q_{0},\vec{q})={N_{c} \over {(2\pi)^3}} {\int_{0}}^{\Lambda} 
{d^3 p \over E_{p} E_{k}}
\left[ (n_{k} - n_{p}) \left\{ {1 \over {E_{p}-E_{k}+q_{0}+i\epsilon}}
+ {1 \over {E_{p}-E_{k}-q_{0}-i\epsilon}} \right\} \right.  \nonumber \\
\times (-E_{p}E_{k} + \vec{p}.\vec{k} + M_{1} M_{2}) \nonumber \\
 + (n_{k} + n_{p}-1) \left\{ {1 \over {E_{p}+E_{k}+q_{0}+i\epsilon}}
+ {1\over {E_{p}+E_{k}-q_{0}-i\epsilon}}\right\}   \nonumber \\
\left. \times (E_{p}E_{k} + \vec{p}.\vec{k} + M_{1} M_{2}) \right]
\label{eq:polar}
\end{eqnarray}
where $N_{c}$ is the number of colours and $\vec{k} = \vec{p} + \vec{q}$.
The energies $E_{p}=\sqrt{p^2+{M_{1}}^2}$ and 
$E_{k}=\sqrt{(\vec{p}+ \vec {q})^2+ {M_{2}}^2}$. 
For pion, $M_{1} = M_{2} = M_{u}$. For kaon, 
$M_{1}= M_{u(d)}$ and $M_{2} = M_{s}$. $n_{k}$ and $n_{p}$ are the 
Fermi-Dirac distribution functions defined earlier.
The pseudoscalar couplings are,
\begin{eqnarray}
G_{\pi}=g_{s}~~+~~g_{D} \gamma \nonumber \\
G_{K^{\pm}}=g_{s}~~+~~g_{D} \beta \nonumber \\
G_{K^{0}}=g_{s}~~+~~g_{D} \alpha
\end{eqnarray}
where $\alpha$, $\beta$ and $\gamma$ are defined in eq. (\ref{eq:defalpha}).

The decay width is evaluated using the imaginary part of the 
eq.(\ref{eq:polar}) as given below,
\begin{eqnarray}
\Gamma_{\phi}=-{{G_{\phi q}}^2 Im\Pi(\omega,\vec{q}\rightarrow 0) \over 
\omega}
\label{eq:decnjl}
\end{eqnarray}
where $G_{\phi q}$ is the empirical meson-quark coupling as obtained in NJL.
Here we have used $G_{\pi q} = 3.5$ and $G_{K q} = 3.6$\cite{n}.

\begin{figure}[b!]
\epsfig{file=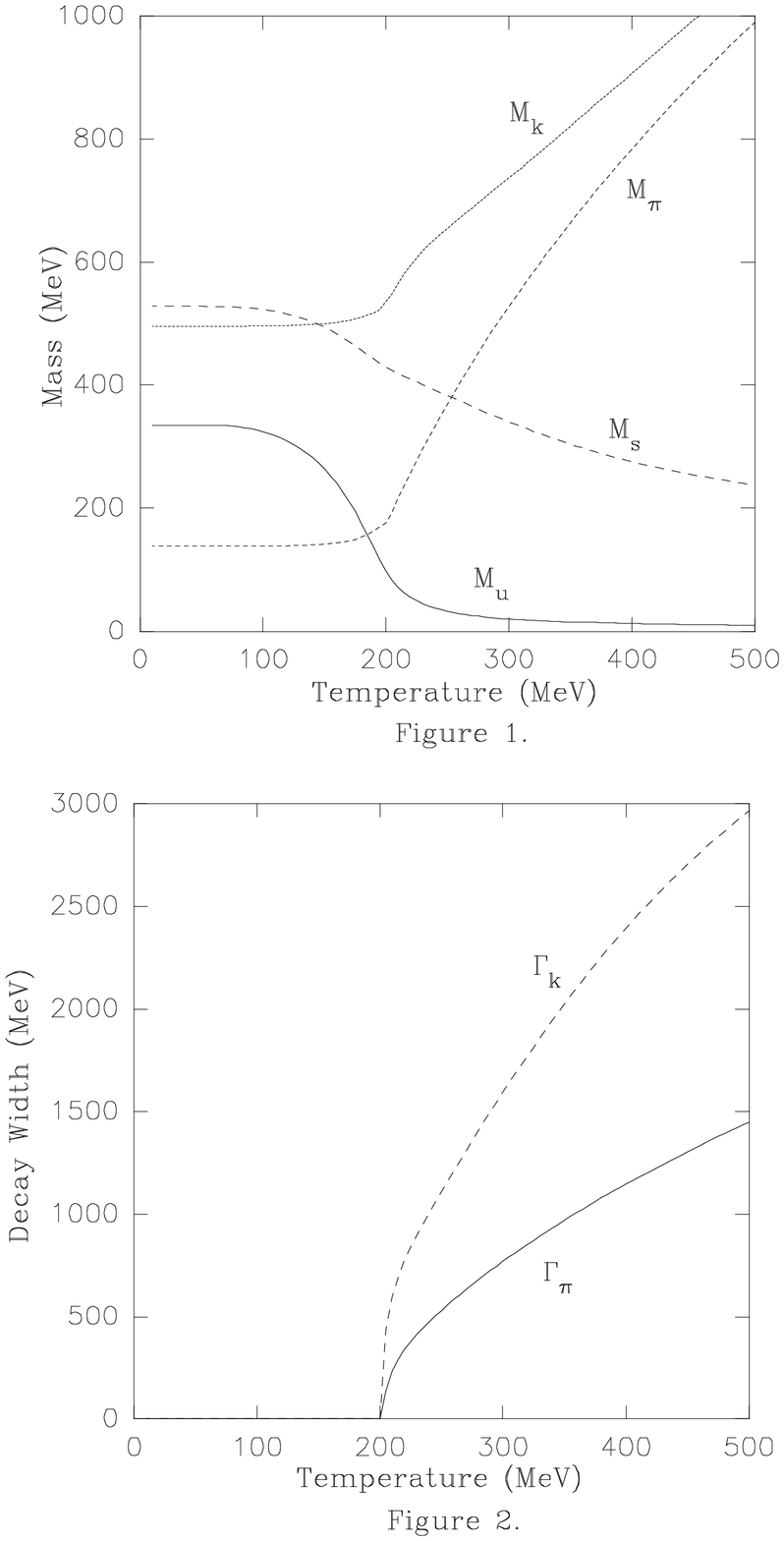,width=6in,height=4in}
\vskip 0.2in
\begin{center}
Figure 1 : Variation of quark and meson masses with temperature.\\
Figure 2 : Variation of meson decay widths with temperature.
\end{center}
\end{figure}

The variation of quark and meson masses is shown in figure 1. The
$u$ or $d$ quark masses starting from 135 MeV drops to the current quark mass
value just after a temperature of 200 MeV. On the other hand, the drop in
the strange quark mass is much smaller around that temperature, showing the 
effect of explicitly broken chiral symmetry, by a larger amount, in the SU(3) 
sector. Pion and kaon both show a similar qualitative behaviour. The masses
of pion and kaon remain constant at their free masses upto a temperature of
200 MeV but increases sharply after that, the pion mass rising to 900 MeV
and kaon mass to 1000 MeV around a temperature of 450 MeV, thus giving a
slower increment for kaons compared to pions.The difference in the behaviour
of kaon and pion  can be attributed to the difference in $u$ and $s$ quark
masses.

Figure 2 shows that the decay width, for both pion and kaon, is very high at
high temperature and decreases with decreasing temperature, going to zero 
at around $T = 0.2$ GeV. It is worth noticing that at around the same 
temperature, the effective pion mass attains the value of the free pion 
mass (figure 1). The decay width of kaon is around 3 GeV where as that of 
pion is around 1.4 GeV at T = 500 MeV. This is  very significant for two
reasons. Firstly, our results show that though there will be pions and kaons
along with the quarks at high temperature phase, the numbers of mesons will
be very small due to their large decay width. Moreover, the number of kaons
will be much less compared to pions at high temperature phase, though both
the mesons will become stable around the same temperature (below 200 MeV).
\par
We thus find that even without any consideration of the detailed evolution
and dynamics of the system, the mesonic modes in a hot quark medium are
found to become important around a temperature of $200$ MeV. Though 
the question whether this is a signature of a phase transition cannot be
addressed within the present framework, the fact that most of the pions and
kaons decay into quarks, owing to a large decay width at temperatures higher
than $T=200$ MeV, is a remarkable finding. Moreover, both the pionic as well 
as kaonic modes start becoming important at about the same temperature, thus
providing a hint of some kind of a transition temperature. It is therefore
tempting to attempt a microscopic investigation of the process of
hadronization within the NJL model.
\par
Recently there have been some attempts to study the formation of hadrons
in quark matter using different semi-microscopic approaches. These studies 
can be characterized either as model dependent calculations \cite{reberg},
or the computer codes based on the string phenomenology \cite{string}, or
other phenomenological description of hadronization \cite{barz}. None
of these approaches account for the essential lack of equilibrium in the
quark-gluon phase. Some efforts have also been made to estimate hadronization
within the parton cascade model by introducing a cut-off to mimic the 
non-perturbative effects \cite{ellis}. To the best of our knowledge, the
first study aimed at investigating the dynamical process of hadronization
in a non-equilibrated quark-gluon system from a physically transparent
approach was in \cite{kyoto}.  
\par
We have earlier shown \cite{g} that perturbative estimates of the 
gluon-gluon, quark-gluon and quark-quark cross sections allow us to 
study the evolution of the quark-gluon matter formed in ultrarelativistic
heavy ion collisions by visualizing the quarks as Brownian particles in a 
hot gluonic thermal bath. Let us start with such a premise, which,
in our opinion, describes the non-equilibrium aspects of the evolution in a
physically transparent manner.
\par
Hadronization in such a system can be studied in the light of Smoluchowski's
theory of coagulation in colloids, which was elaborated further by
Chandrasekhar \cite{chandra}. This theory suggests that coagulation 
results as a consequence of each colloidal particle being surrounded by
a sphere of influence of a certain radius $R$ such that the Brownian 
motion of a particle proceeds unaffected only so long as no other particle
comes within its sphere of influence. When another Brownian particle does 
come within a distance $R$ of the test particle, they form a two body cluster. 
This cluster also describes a Brownian motion but at a reduced rate due to 
its increased size/mass. The process continues further till a single cluster 
of all the particles is formed. 
\par 
In order for this stochastic scenario of cluster formation to apply 
to a system of Brownian quarks in the 
hot gluon bath, it is essential that each quark has an appropriate sphere of 
influence of radius $r$. Obviously, this radius $r$ will depend on the 
spin-isospin combination of final cluster (whether the final cluster is 
scalar, pseudoscalar, vector or axial vector meson or even a baryon).
Mesons are formed when one quark and one antiquark with proper quantum numbers
come within the spheres of influence of each other. (Clusters with 
greater numbers of quarks and antiquarks (e.g. baryons) can also be formed by 
imposing the conditions of colour neutrality and charge balance properly.) 
This implies that the radius of the sphere of influence corresponds to the 
correlation length between the quarks in the proper hadronic channel. In 
other words, this is the screening length of the corresponding hadrons in 
the hot quark-gluon matter.  (One can immediately see, without further ado,
that the rate of pion formation in the hot quark matter should be rather
small at high temperatures and increase with falling temperature as the pion
screening length (inverse of the screening mass) decreases with increasing
temperature in all realistic pictures.) The pions formed at very high
temperatures are most likely to decay back into quarks and antiquarks,
because of the large decay width at high temperatures. We adopt, for the
sake of consistency, the NJL model estimates for the pion screening mass
\cite{flor} in the present work. 
\par
The rate of pion formation from Brownian quarks as a stochastic 
process, as is evident from the preceding discussion, 
depends on the number of quarks (antiquarks) falling into the sphere of
influence of another antiquark (quark), the number of pions decaying back to 
quarks and antiquarks and also the change in the pion density due to the 
expansion of the system. Thus the rate equations can be written as,
\begin{eqnarray}
{dn_{\pi^{a}} \over dt}= n_{q_i} n_{\bar{q}_j} 4\pi r^2<\vec{v}\cdot \hat{r}>
- {\Gamma^{total}}_{\pi^{a} \rightarrow q_{i} {\bar q}_{j}} - {n_{\pi^a}\over t}  
 \label{eq:dp1dt} \\
{dn_{\pi^{0}} \over dt}= {1\over 2} (n_{u} n_{\bar{u}}+ n_{d} n_{\bar{d}})     
4\pi r^2<\vec{v}\cdot \hat{r}> 
-{\Gamma^{total}}_{{\pi^0} \rightarrow u \bar{u}
(d{\bar d})} - {n_{\pi^0}\over t}  {\label{eq:dp2dt}}\\
{dn_{q_{i}} \over dt} =  {\Gamma^{total}}_{g\rightarrow q_{i} \bar{q}_{i}}
+{\Gamma^{total}}_{gg\rightarrow q_{i} \bar{q}_{i}}    
- n_{q_i} n_{\bar{q}_j} 4\pi r^2 <\vec{v}\cdot \hat{r}>+ 
{\Gamma^{total}}_{\pi \rightarrow q_{i} {\bar q}_{j}} - {n_{q_i}\over t}
{\label{eq:dqdt}}  
\end{eqnarray}

In eqs.(\ref{eq:dp1dt},\ref{eq:dp2dt}), the first term is the rate of 
pion formation 
($a\equiv$ + or -); the second term is the rate of pions decaying back to 
quarks and the third term is due to  Bjorken (longitudinal) expansion of 
the system. $i(j)$ stands for $u$ or $d$ (we ignore $s$ and other heavier 
flavours). $<\vec {v} \cdot \hat{r}>$ (the average 
relative velocity in the radial direction) is calculated using the 
J\"uttner distribution,
\begin{equation}
f(x,p)=e^{-\beta p \cdot u(x)} \label{eq:jutt}
\end{equation}
There would also be a corresponding rate equation for the antiquarks, which
looks exactly like eq. (\ref{eq:dqdt}) and hence not explicitly written.
In eq. (\ref{eq:dqdt}) the ${\Gamma^{total}}_{g\rightarrow q \bar q}$ as well
as ${\Gamma^{total}}_{gg\rightarrow q \bar q}$ stand for the corresponding
net quantities.
\par
As mentioned earlier, we are considering a non-equilibrated quark matter
and hence the pions formed will also be out of equilibrium. This is
taken into account by multiplying the relevant distribution functions with 
the  ratios $r_{q}=n_{q}/n_{eq}$ and $r_{\pi}=n_{\pi}/n_{e\pi}$ where
$n_{q}$ and $n_{eq}$ are non-equilibrium and equilibrium densities of
quarks and $n_{\pi}$ and $n_{e\pi}$ are non-equilibrium and equilibrium
densities for pions. 
\par
In all these expressions, the appropriate masses are the effective masses 
including the current as well as thermal contribution,
, whose importance in determining the dynamics of the hot quark matter
has been well established. For quarks (antiquarks), this is 
$$m_{eff}=\sqrt{{m_{q}(curr)}^{2}+{m_{q}(thermal)}^{2}}$$ 
where \cite{alther,thoma}
\begin{equation}
{m_{q}^{2}}(thermal)= (1+{r_{q}\over {2}}) ({{g_{s}T} \over 3})^{2}
\end{equation}
and $m_{q}(curr)$ is taken to be $10$ MeV. For gluons the thermal mass is,
\begin{equation}
m_{g}(thermal)={2\over 3} g_{s}T
\end{equation}
The running coupling constant $\alpha_{s}$ as a function of temperature is 
given by \cite{geiger}
\begin{equation}
\alpha_{s}={{12\pi} \over {(33-2n_{f})ln \left[ {\bar{Q^2} \over \Lambda^2} \right]}}
\end{equation}
with $\bar{Q^2} = {m_{eff}}^{2}(T) + 9T^{2}$. 
\par
Simultaneously, we must take account of energy momentum conservation
which, for a Bjorken flow, corresponds to the following equation
\begin{equation}
{\partial{\epsilon} \over {\partial t}}= - {{\epsilon + P} \over t}
\label{eq:energy}
\end{equation}
where $\epsilon\equiv \epsilon_{total} = \epsilon_{g} + \epsilon_{q}
+ \epsilon_{\pi}$. We also include the one loop correction to
$\epsilon_{g}$ \cite{plumer}. $\epsilon$ and $P$ are related through
the velocity of sound, as in \cite{g}. For a complete description of 
the system, eqs. ({\ref{eq:dp1dt}), (\ref{eq:dp2dt}), (\ref{eq:dqdt}) 
and (\ref{eq:energy}) must be solved self-consistently. The initial 
conditions are taken from \cite{g} for RHIC energies. The initial 
time ($t_{g}$) is the time when gluons thermalise (=0.3 fm), where 
$r_{q}$ = 0.15, $r_{g}$=1 and $r_{\pi}$ is taken to be 0. The 
temperature at this time is 500 MeV. Note that we are working at $y=0$ 
so that $t$ and $\tau$ are the same and the baryon chemical potential is
zero.
\par
\begin{figure}[htb]
\epsfig{file=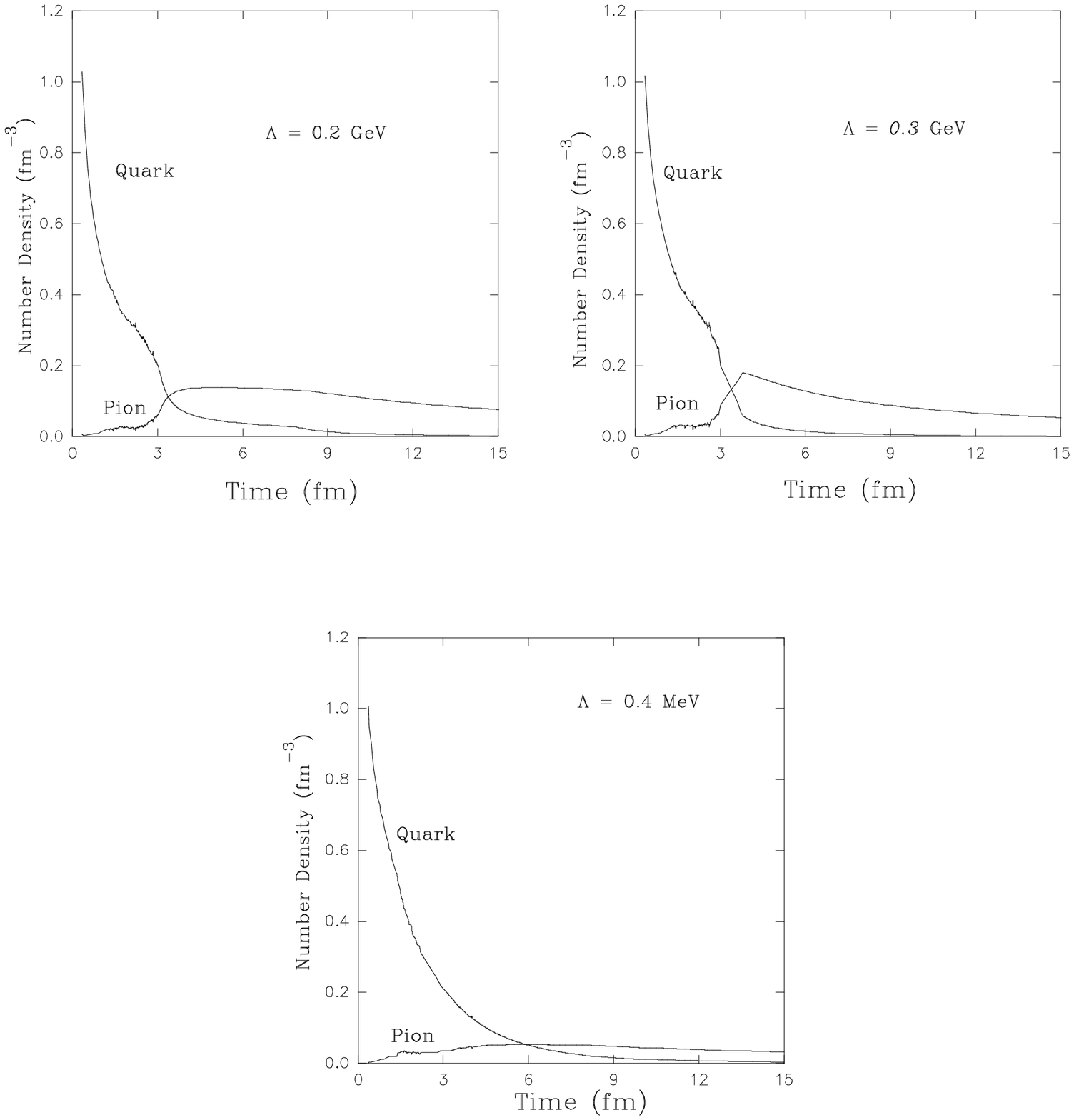,width=6in,height=3in}
\vskip 10pt
\begin{center}
Figure 3 : Time evolution of quark and pion number densities for various
values of QCD parameter $\Lambda$.
\end{center}
\end{figure}
\par 
Figure 3 shows the variation of pion and quark number densities with 
time, for various values of the QCD parameter $\Lambda$. In all three cases, 
we find the same qualitative feature that pions start appearing in the system
quite early on but they become appreciable in number only after some time.
At late times the system is dominated by pions. This cross over occurs 
at $t \geq $ 4 fm for $\Lambda$ = 0.2 or 0.3 GeV while for $\Lambda$=
0.4 GeV this happens at $t\sim$ 6 fm. 
\par
Figure 4 shows the variation of temperature with time. Obviously, there is
a dramatic effect of the QCD parameter $\Lambda$. In all the cases, there
is a change at $T\sim$ 215 MeV, corresponding to $t\sim$ 3.5 fm; the 
variation of temperature with time becomes slower, as is expected in the
mixed phase of a first order phase transition. At $\Lambda$=0.2 GeV,
this occurs for a very short period of time, before the system starts cooling
again. The duration of the constant temperature configuration  increases
with $\Lambda$, and for $\Lambda$=0.4 GeV, it persist upto 9 fm before the
temperature of the system starts falling again. 
\par
\begin{figure}[htb]
\epsfig{file=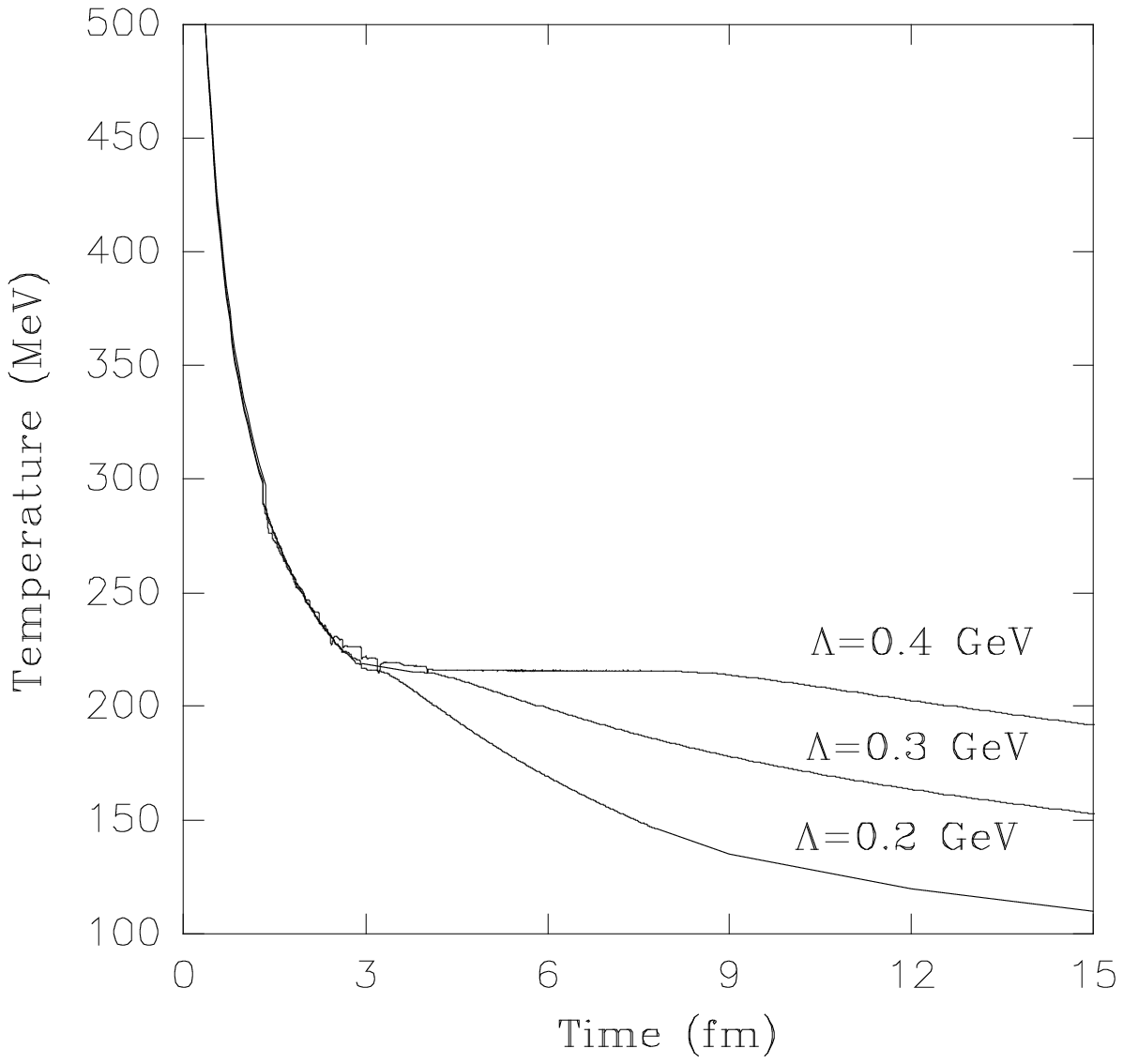,width=6in,height=2.5in}
\vskip 10pt
\begin{center}
Figure 4 : Time evolution of the Temperature for various values of 
QCD parameter $\Lambda$.
\end{center}
\end{figure}
\par
Obviously, this is a clear indication of an {\it apparent} first order
transition. Microscopically, the appearance of the mixed phase at
a temperature of $\sim$ 215 MeV can be understood from the fact that
the pion decay width goes to zero at such a temperature.
All the pions that were formed earlier in the system tended to decay back
to quarks and antiquarks on a fast time scale. Only after the pion decay
width becomes small would the formed pions become stable. 
\par
The apparently desirable features of the dynamical deconfining transition
seem to arise quite naturally in this scenario. It is therefore important
to test if all the conservation laws are obeyed during the evolution. To
this end, let us check if entropy increases steadily during the entire
evolution.
\par
\begin{figure}[htb]
\epsfig{file=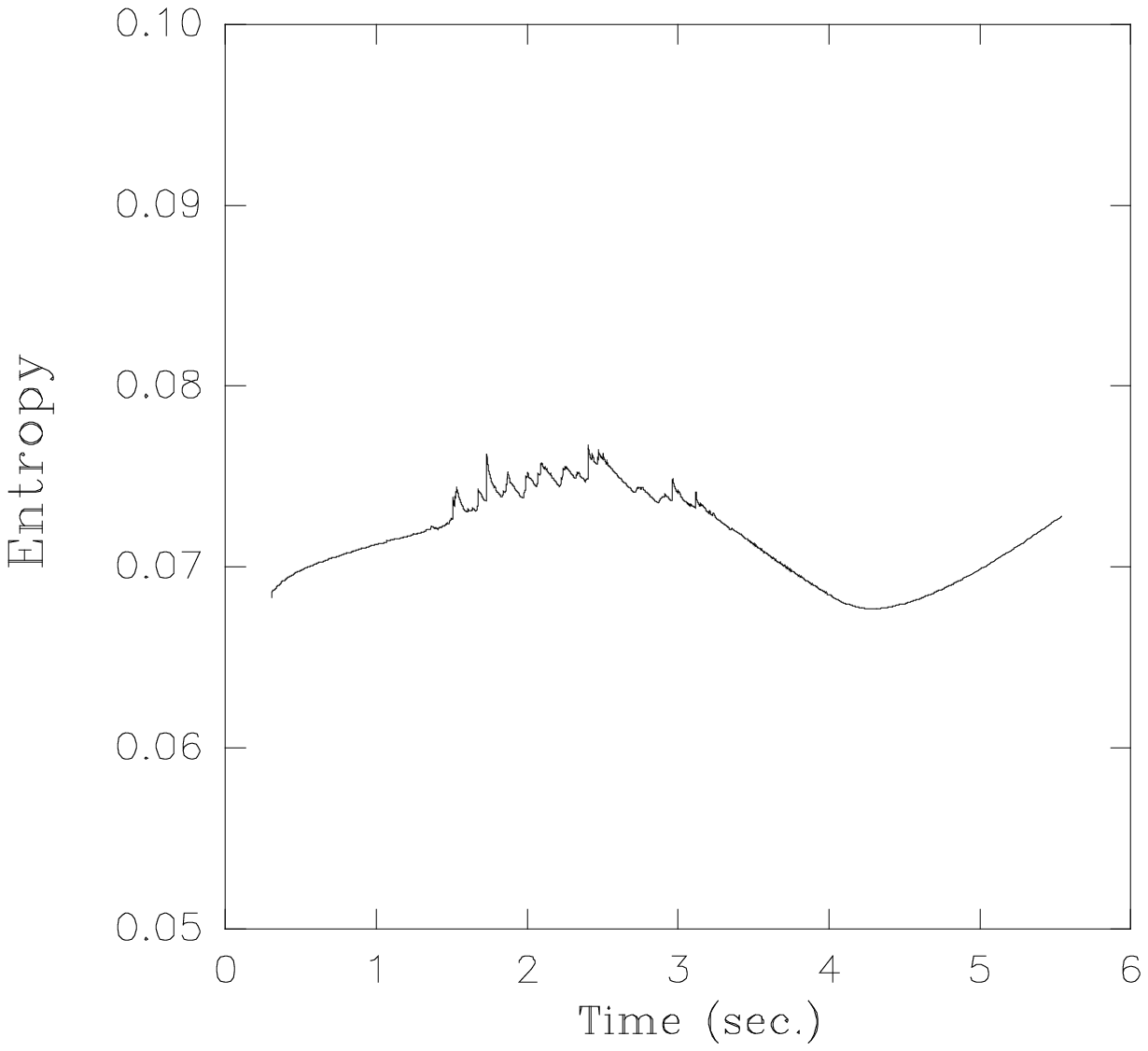,width=6in,height=2.5in}
\vskip 10pt
\begin{center}
Figure 5 : Time evolution of entropy with time ($\Lambda$ = 0.3 GeV).
\end{center}
\end{figure}
\par
The decrease in entropy occurs precisely where the gluons drop out of the
system. This, in hindsight, was to be expected, since the
sudden decrease in the number of degrees of freedom in the system should
lead to a decrease in entropy. The NJL model, in the present form, does not
account for the gluonic degrees of freedom. It thus appears to us that in
order to obtain an effective low energy lagrangian from QCD capable of
describing the dynamical evolution of the quark-gluon system, the role of
the gluons must be taken into account. Incorporation of the dilaton field
to account for the QCD trace anomaly in the NJL model \cite{Jaminon} is
indeed a promising step in this respect, but the entropy carried by the
gluons at high temperature should also be accommodated. Such a study is on
our current agenda.
\par
It is a great pleasure to thank the organisers of the workshop, and
especially Prof. J. da Providencia, for the kind invitation and their warm
hospitality. This talk is based on a continuing collaboration the author has
had with Jan-e Alam, Abhijit Bhattacharyya, Sanjay Kumar Ghosh and Bikash
Sinha over the past several years. I take this opportunity to thank them all.

\end{document}